\documentclass[aps,prl,twocolumn,superscriptaddress, longbibliography]{revtex4-1}
\usepackage{graphicx}
\usepackage{latexsym}
\usepackage{amssymb}
\usepackage{amsmath}
\usepackage{amsfonts}
\usepackage{upgreek}
\usepackage{subfigure}
\usepackage{bm}
\usepackage{nicefrac}
\usepackage{bbold}
\usepackage{verbatim}
\usepackage[unicode=true,
 bookmarks=false,
 breaklinks=false,pdfborder={0 0 1},backref=false,colorlinks=true]
 {hyperref}
\hypersetup{
 linkcolor=[rgb]{0,0,1},citecolor=[rgb]{0,0,1},urlcolor=[rgb]{0,0,1}}
\usepackage{multirow}
\usepackage{color}
\usepackage{comment}
\usepackage{xcolor}
\usepackage{soul}
\usepackage{tikz}

\usepackage{braket}

\begin{document}

\title{Kinetic-to-magnetic frustration crossover and linear confinement \\ in the doped triangular $\mathbf{t-J}$ model}

\author{Henning Schl\"omer}
\affiliation{Department of Physics and Arnold Sommerfeld Center for Theoretical Physics (ASC), Ludwig-Maximilians-Universit\"at M\"unchen, Theresienstr. 37, M\"unchen D-80333, Germany}
\affiliation{Munich Center for Quantum Science and Technology (MCQST), Schellingstr. 4, D-80799 M\"unchen, Germany}
\affiliation{ITAMP, Harvard-Smithsonian Center for Astrophysics, Cambridge, MA, USA}
\affiliation{Department of Physics and Astronomy, Rice University, Houston, Texas 77005, USA}
\author{Ulrich Schollw\"ock}
\affiliation{Department of Physics and Arnold Sommerfeld Center for Theoretical Physics (ASC), Ludwig-Maximilians-Universit\"at M\"unchen, Theresienstr. 37, M\"unchen D-80333, Germany}
\affiliation{Munich Center for Quantum Science and Technology (MCQST), Schellingstr. 4, D-80799 M\"unchen, Germany}
\author{Annabelle Bohrdt}
\affiliation{Munich Center for Quantum Science and Technology (MCQST), Schellingstr. 4, D-80799 M\"unchen, Germany}
\affiliation{ITAMP, Harvard-Smithsonian Center for Astrophysics, Cambridge, MA, USA}
\affiliation{Department of Physics, Harvard University, Cambridge, Massachusetts 02138, USA}
\affiliation{Institut für Theoretische Physik, Universität Regensburg, D-93035 Regensburg, Germany}
\author{Fabian Grusdt}
\affiliation{Department of Physics and Arnold Sommerfeld Center for Theoretical Physics (ASC), Ludwig-Maximilians-Universit\"at M\"unchen, Theresienstr. 37, M\"unchen D-80333, Germany}
\affiliation{Munich Center for Quantum Science and Technology (MCQST), Schellingstr. 4, D-80799 M\"unchen, Germany}

\date{\today}
\begin{abstract}
Microscopically understanding competing orders in strongly correlated systems is a key challenge in modern quantum many-body physics. For example, the study of magnetic polarons and their relation to pairing in the Fermi-Hubbard model in different geometries remains one of the central questions, and may help to understand the mechanism underlying unconventional superconductivity in cuprates or transition metal dichalcogenides. With recent advances in analog quantum simulation of the Fermi-Hubbard model on non-bipartite lattices, frustrated physics can now be explored experimentally in systems lacking particle-hole symmetry. Here, we study the singly doped $t-J$ model on the triangular lattice, focusing on the competition between kinetic and magnetic frustration as a function of temperature. In doublon doped systems, we uncover a crossover between Nagaoka-type ferromagnetic (FM) correlations at high temperature and exchange mediated antiferromagnetic (AFM) order around the doublon at low temperature. For hole doped systems, kinetic Haerter-Shastry-type AFM at high temperature as well as exchange interactions at low temperature favor $120^{\circ}$ order, strengthening magnetic correlations compared to the undoped system. In the ground state, the presence of AFM correlations throughout a wide range of interactions indicates confinement of both types of dopants. In this regime we firmly establish the presence of linear confining potentials via energy scaling arguments, supporting the picture of geometric strings in the frustrated triangular $t-J$ model. 
\end{abstract}
\maketitle

\textit{Introduction.---}  The study of magnetic polarons in doped Mott insulators~\cite{Sachdev1989, KLR1989, Koepsell_nature2019, Miyamoto2018, Grusdt_strings, Shon2019, GrusdtX, Blomquist2020, Ji2021} is at the heart of strongly correlated materials. They feature the same interplay of magnetism and motional degrees of freedom which is believed to give rise to phenomena such as the pseudogap and high-temperature superconductivity, rendering a thorough comprehension of their competition an essential step towards developing microscopic theories~\cite{Lee2006, Keimer2015}.

\begin{figure}
\centering
\includegraphics[width=0.95\columnwidth]{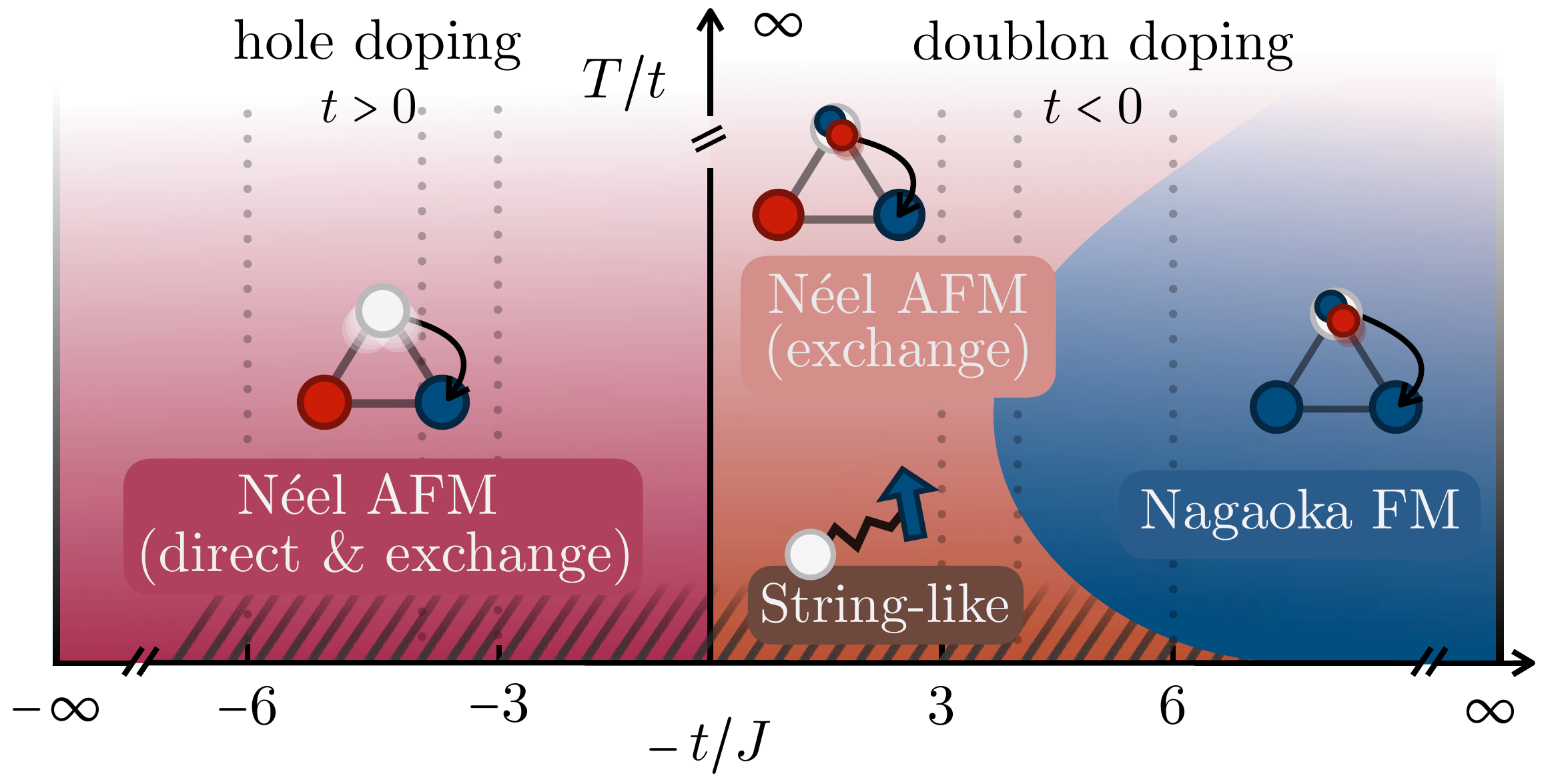}
\caption{Schematic phase diagram of the triangular $t-J$ model, indicating different regimes of itinerant magnetism for a single dopant. For hole doping, both direct and exchange mechanisms favor $120^{\circ}$ N\'eel order (light red area). On the doublon doped side, Nagaoka FM and kinetic frustration compete, leading to AFM (FM) spin correlations around the dopant in the orange (blue) regions. Dashed lines illustrate parameters of our numerical simulations. At low temperatures, singly doublon doped systems are well described by geometric strings (grey hatched area; for holes at low temperatures the signatures of strings are less striking). At high temperatures, small AFM correlations are present for all values of $t/J$, leading to reentrance phenomena between AFM and FM correlations on the doublon doped side.}
\label{fig:pd}
\end{figure}

\begin{figure*}
\centering
\includegraphics[width=\textwidth]{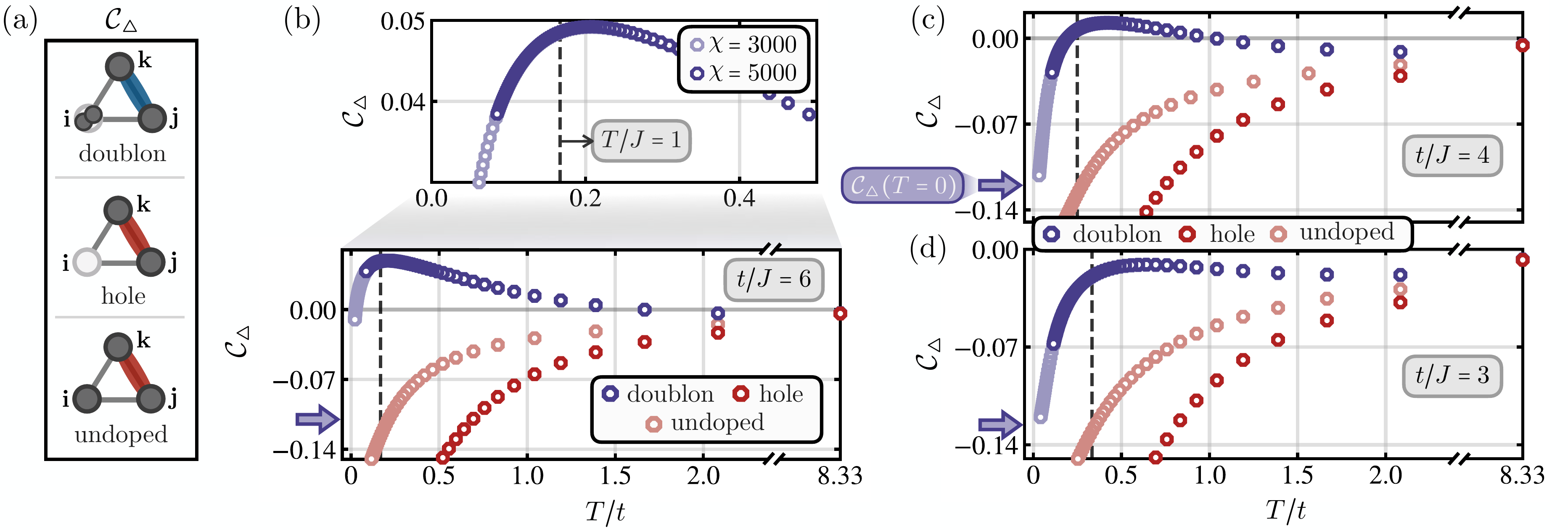}
\caption{(a) We compute thermally averaged charge-spin-spin correlations around the dopant, Eq.~\eqref{eq:hss_corr}, as a function of temperature $T/t$, and for $\mathbf{i} = [x=5,y=2]$. (b) For $t/J = 6$ and doublon doping (blue data points), Nagaoka-type kinetic FM leads to positive correlations at temperatures $T \sim t$. At $T \sim J$ (grey dashed line), a maximum of correlations is observed as exchange processes start to win over Nagaoka-type FM. Ground state results of $\mathcal{C}_{\triangle}$ for doublon doping are shown by blue arrows on the left. The inset shows the region around the maximum for MPS calculations with bond dimensions $\chi = 3000$ and $\chi = 5000$, demonstrating convergence down to $T/t \sim 0.1$. Two-point correlations in the undoped case are shown in salmon. Upon hole doping the system, both exchange and the release of kinetic frustration favor $120^{\circ}$ order, strengthening AFM correlations (red dots). For $t/J=4$, (c), correlations around the doublon are positive in the range $0.3\lesssim T/t \lesssim 1$, whereas for $t/J = 3$, (d), AFM correlations win over Nagaoka-type FM and $\mathcal{C}_{\triangle}$ is negative for all temperatures. Nevertheless, kinetic effects manifest in the doublon case through a maximum of the correlator at $T/J \sim 1$. We do not show the full range of correlations for the hole doped case for illustrative reasons, and show their convergence behavior down to low temperatures in~\cite{supp_mat}.}
\label{fig:fT}
\end{figure*}

Analog simulations via ultracold atoms yield a highly controllable, pristine platform to experimentally explore paradigmatic Hamiltonians such as the Fermi-Hubbard model~\cite{Esslinger2010, Bohrdt2020}, which is believed to capture the essential physics of doped Mott insulators. On non-bipartite lattices, the notion of frustration and absence of particle-hole symmetry can lead to distinct physical phenomena~\cite{Haerter2005, Merino2006, Sposetti2014, Batista, Morera2021, Zhu2022_2, Zhu2022, Morera2022, Kraats2022, Kadow2022} which, with recent successful implementations of optical triangular lattices, can now be explored experimentally~\cite{Struck2011, Jo2015, Yamamoto2020, Schauss2021_tri, Greiner2022_tri}. Furthermore, solid state platforms based on Moir\'e heterostructures constitute a promising second pillar for quantum simulation of Hubbard physics on triangular lattice geometries~\cite{Wu2018, Tang2020, Kennes2021}, their relation to magnetic polarons being pointed out in~\cite{Davydova2022, Lee2022}.

The delocalization of particles associated with kinetic energy gain is of fundamental importance when describing itinerant magnetism in strongly correlated systems. 
Effective antiferromagnetic (AFM) spin-spin interactions arise in the Hubbard model via virtual exchange (kinetic) processes, which favor N\'eel order in undoped and weakly doped systems~\cite{MultiMess}. In contrast, in the infinitely interacting limit -- where exchange processes are fully suppressed -- Nagaoka's theorem states that, on bipartite lattices and for a single dopant, the spin background of the ground state is fully spin polarized, above which the dopant can minimize its kinetic energy~\cite{Nagaoka1966, Becca2001, White2001}. However, on non-bipartite lattices, destructive interference of dopant paths can lead to the appearance of kinetic frustration, a quantum effect where the mobile dopants are prohibited to minimize their kinetic energy in a fully spin polarized background. This, in turn, leads to the formation of Haerter-Shastry-type classical AFM order via the release of kinetic frustration~\cite{Haerter2005, Sposetti2014, Zhang_triangular_2018, Morera2021, Morera2022}. In the infinitely interacting Hubbard model ($U\rightarrow \infty$) on the triangular lattice, it has been shown that kinetic effects result in both Nagaoka-type FM and Haerter-Shastry-type AFM depending on whether the system is slightly doublon or, respectively, hole doped~\cite{Morera2022}. 

Here, we study the interplay between kinetic and exchange mechanisms for itinerant magnetism in the triangular $t-J$ model for the experimentally relevant regime of finite temperature and interaction strengths, focusing on single hole and doublon doping. For hole doping, exchange and kinetic processes are in synergy with each other, both favoring $120^{\circ}$ N\'eel order in the magnetic background, which strengthens AFM correlations. For doublon doping, we find that the competition between Nagaoka-type FM and Heisenberg AFM gives rise to crossovers between FM and AFM spin correlations around the doublon as a function of temperature. Our results are summarized in Fig.~\ref{fig:pd}.

In the ground state, where magnetic correlations are governed by exchange-mediated Heisenberg interactions, we firmly establish the Brinkman-Rice picture~\cite{Brinkman1970, Shraiman1988} of linear confining potentials arising from the antiferromagnetically ordered spin background (geometric string theory~\cite{GrusdtX, Grusdt_strings, Chiu2019}) for the doped doublon. For the doped hole, our results are also consistent with the string picture, although the signatures we find are less striking. We map out the various regimes using ground state and finite temperature density matrix renormalization group (DMRG) techniques~\cite{SchollwoeckDMRG2, Schollwoeck_DMRG, WhiteDMRG, Paeckel_time} on cylinders of the $t-J$ model.

\textit{The model.---} We consider the $t-J$ Hamiltonian on the triangular lattice, 
\begin{equation}
\begin{aligned}
    \hat{\mathcal{H}} = -t \sum_{\sigma, \braket{\mathbf{i}, \mathbf{j}}} \hat{\mathcal{P}}_{GW} &\big(\hat{c}_{\mathbf{i}, \sigma}^{\dagger} \hat{c}_{\mathbf{j}, \sigma} + \text{h.c.} \big)\hat{\mathcal{P}}_{GW} \, + \\ & + J \sum_{\braket{\mathbf{i}, \mathbf{j}}} \left( \hat{\mathbf{S}}_{\mathbf{i}} \cdot \hat{\mathbf{S}}_{\mathbf{j}} - \frac{\hat{n}_{\mathbf{i}}\hat{n}_{\mathbf{j}}}{4} \right),
\end{aligned}
\label{eq:tri_tJ_H}
\end{equation}
where $\hat{c}_{\mathbf{i}, \sigma}^{(\dagger)}$, $\hat{n}_{\mathbf{i}}$ and $\hat{\mathbf{S}}_{\mathbf{i}}$ are fermionic annihilation (creation), particle density, and spin operators on site $\mathbf{i}$, respectively; $\braket{\mathbf{i}, \mathbf{j}}$ denotes nearest neighbor (NN) sites on the 2D triangular lattice, and $\hat{\mathcal{P}}_{GW}$ is the Gutzwiller operator projecting out states with double occupancy. In the following, we fix the total particle number to $N = N_L - 1$, with $N_L$ the number of lattice sites. For $t>0$, Eq.~\eqref{eq:tri_tJ_H} then describes a single hole hopping through a Mott insulator on the triangular lattice. Doublon doping, on the other hand, is captured by Eq.~\eqref{eq:tri_tJ_H} still at $N = N_L - 1$ but with $t<0$, which can be seen from a particle-hole transformation of the $t-J$ Hamiltonian~\cite{Zheng2022}. From now on, we will denote the hopping strength as $t$ for both hole and doublon doped systems, and implicitly imply its corresponding sign in Eq.~\eqref{eq:tri_tJ_H}.

\textit{Finite temperature correlations.---} We simulate the doped triangular $t-J$ model, Eq.~\eqref{eq:tri_tJ_H}, at finite temperature using imaginary time evolution schemes (purification~\cite{Nocera2016}) via matrix product states (MPS)~\cite{SchollwoeckDMRG2, Schollwoeck_DMRG, WhiteDMRG, Paeckel_time}. 
We implement the $\mathrm{U}(1)$ particle conservation symmetry, but allow for thermal spin fluctuations~\cite{supp_mat}. We simulate a cylinder of size $L_x \times L_y = 9 \times 3$, and apply open (periodic) boundaries along the $x-$ ($y-$) direction, which can resolve $120^{\circ}$ N\'eel order.  

We calculate the thermal average at temperature $T$, \begin{equation} \mathcal{C}_{\triangle} = \braket{\hat{P}^h_{\mathbf{i}} \hat{\mathbf{S}}_{\mathbf{j}} \cdot \hat{\mathbf{S}}_{\mathbf{k}}}_T/\braket{\hat{P}^h_{\mathbf{i}}}_T,
\label{eq:hss_corr}
\end{equation} where $\hat{P}^h_{\mathbf{i}}$ is a local projection onto the hole (dopant) at site $\mathbf{i}$, i.e., $\hat{P}^h_{\mathbf{i}} = \ket{0}_{\mathbf{i}} \bra{0}_{\mathbf{i}}$. We choose indices $(\mathbf{i,j,k})$ to lie on a triangular plaquette in the bulk of the system, i.e., we focus on spin-spin correlations in the direct vicinity of the dopant, see Fig.~\ref{fig:fT}~(a). To compare to the undoped case, we further calculate finite temperature two-point correlations of the Heisenberg model on the triangular lattice, where $\mathcal{C}_{\triangle}$ reduces to $\braket{\hat{\mathbf{S}}_{\mathbf{j}} \cdot \hat{\mathbf{S}}_{\mathbf{k}}}_T$. Full spatial and spin resolution techniques in quantum gas microscopy allow for a direct access of spin correlations around the mobile dopant~\cite{Boll2016, Koepsell2020, Bakr_microscope}.

Results showing $\mathcal{C}_{\triangle}(T)$ for $t/J = 6, 4, 3$ are presented in Fig.~\ref{fig:fT}~(b)-(d). Two-point correlations in the undoped Heisenberg model are shown in each case with salmon dots, where Heisenberg interactions cause AFM correlations to build up as the temperature is lowered. In Fig.~\ref{fig:fT}~(b), $t/J = 6$ is simulated. When doping a hole into the system, both direct ($t$) and indirect ($J$) kinetic mechanisms are in synergy with each other, i.e., $120^{\circ}$ N\'eel order is favored. In particular, stronger AFM signals are visible in the hole doped system compared to its undoped analog, shown by red data points in Fig.~\ref{fig:fT}.

When instead doping a single doublon into the system, Nagaoka-type kinetic FM starts to develop and overpower weak AFM correlations at $T/t \lesssim 1$. FM correlations increase with decreasing temperature, until they reach a maximum at $T/J \lesssim 1$, where Heisenberg interactions favoring AFM alignment visibly suppress positive correlations. In fact, in the ground state, correlations around the hole are negative, as indicated by the blue arrow in Fig.~\ref{fig:fT}~(b). The zoom-in panel in Fig.~\ref{fig:fT}~(b) shows the peak region for TDVP-2 calculations with maximal bond dimensions $\chi = 3000$ and $\chi = 5000$. We observe that the peak is well converged for the two given bond dimensions down to $T/t \sim 0.1$. We note that the sharp drop of the correlator $\mathcal{C}_{\triangle}$ towards the ground state value at low temperatures is particularly hard to resolve using purification. Nevertheless, energies are found to converge towards the ground state results at low temperatures~\cite{supp_mat}. 

The competition between Nagaoka-type kinetic FM and Heisenberg AFM on the doublon doped side is underlined in Fig.~\ref{fig:fT}~(c) for $t/J = 4$. Here, next to an apparent maximum at $T/J \sim 1$ of $\mathcal{C}_{\triangle}$, a sign change of these charge-spin-spin correlations around the doublon is observed at $T/J < 1$. Small residual negative correlations at high temperatures, positive correlations at $T/t\sim 1$ and negative correlations at low temperatures hence lead to reentrance phenomena of the Nagaoka-type regime as a function of temperature, see Figs.~\ref{fig:pd} and~\ref{fig:fT}. For $t/J=3$, correlations are observed to be negative throughout all temperatures, as presented in Fig.~\ref{fig:fT}~(d). Nevertheless, kinetic effects and their interplay with Heisenberg-type AFM are clearly visible -- i.e., AFM correlations around a doublon are weakened and show a maximum at $T/J \sim 1$.

\textit{Linear confinement.---} We now turn to the ground state of the system. 
Using DMRG, we simulate Eq.~\eqref{eq:tri_tJ_H} on a cylinder and employ $U(1)$ symmetries both in particle number and total magnetization. We focus on system sizes $L_x \times L_y = 12 \times 6$ with open (periodic) boundaries along the $x-$ ($y-$) direction, and use bond dimensions of $\chi = 5500$.

\begin{figure}
\centering
\includegraphics[width=0.84\columnwidth]{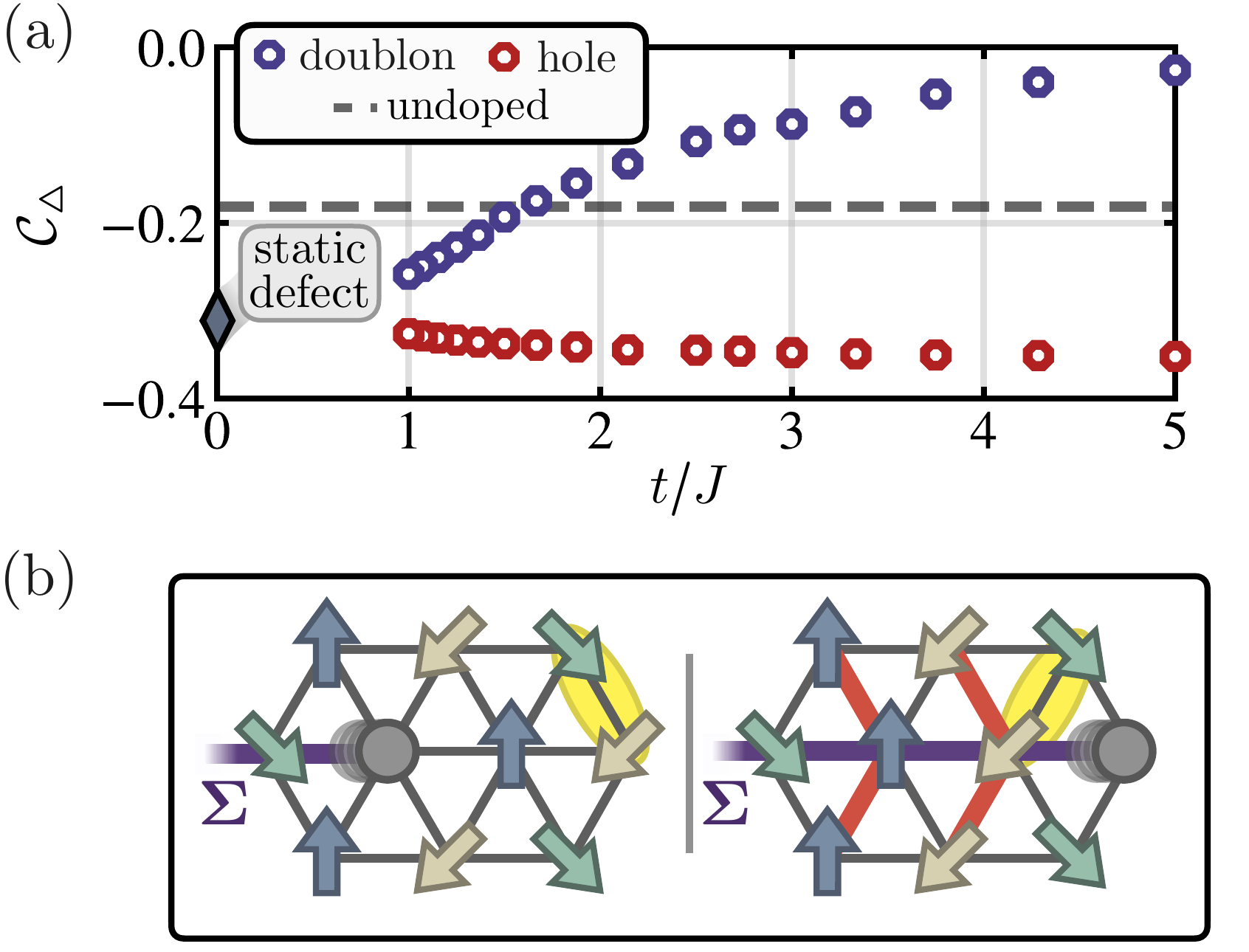}
\caption{(a) Ground state spin-spin correlations around the dopant, $\mathcal{C}_{\triangle}$, as a function of $t/J$ for a doped hole (red) and doublon (blue), evaluated for $\mathbf{i} = [x=4,y=3]$. Two-point correlations in the triangular Heisenberg model are shown by the grey dashed line; the blue diamond shows $\mathcal{C}_{\triangle}$ surrounding an immobile dopant ($t/J\rightarrow 0$). (b) AFM correlations around the dopant in the ground state suggests a linear confining potential imposed by the spin background. When the dopant hops through the system, it displaces the spins, leading to a linearly growing energy penalty (red lines) as a function of the string length $|\Sigma|$ (purple line)~\cite{Grusdt_strings, GrusdtX} and strong AFM correlations around the dopant (indicated by yellow ellipses).}
\label{fig:gs_corr}
\end{figure}

Fig.~\ref{fig:gs_corr}~(a) shows $\mathcal{C}_{\triangle}$ as a function of $t/J$. In the hole doped case, AFM correlations are significantly stronger compared to corresponding two-point spin correlations in the undoped case (grey dashed line), with a subtle tendency towards weaker correlations for decreasing $t/J$. On the doublon doped side (and away from the Nagaoka regime at $t/J \gg 1$), strong AFM correlations quickly develop for decreasing $t/J$, complementing the results at finite temperature. In the limit $t/J \rightarrow 0$, i.e. an immobile defect doped into a Mott insulator, surrounding AFM correlations are expected to be stronger compared to the undoped case due to a local reduction of frustration around the defect~\cite{Kraats2022}, see the dark blue diamond in Fig.~\ref{fig:gs_corr}~(a).

The observed strong AFM correlations in the ground state for both hole and doublon doping and for a large range of $t/J$ suggest linear confinement of the dopant, illustrated in Fig.~\ref{fig:gs_corr}~(b). In particular, when hopping through a $120^{\circ}$ ordered, frozen spin background, each hopping process causes a magnetic energy cost of order $J$ and creates a (partial) memory of the chosen trajectory. This is expected to lead to the formation of a geometric string $\Sigma$ along which pairs of spins are locally aligned~\cite{Brinkman1970, GrusdtX, Grusdt_strings}. A hallmark of such string formation is the resulting scaling of the dopant's energy when $J/t \ll 1$ as~\cite{Brinkman1970, Shraiman1988}
\begin{equation}
    E/t = -2\sqrt{z-1} + const. \, (J/t)^{2/3} + \mathcal{O}(J/t),
    \label{eq:gst_E_scaling}
\end{equation}
with $z$ the connectivity of the underlying lattice. 
\begin{figure}
\centering
\includegraphics[width=0.93\columnwidth]{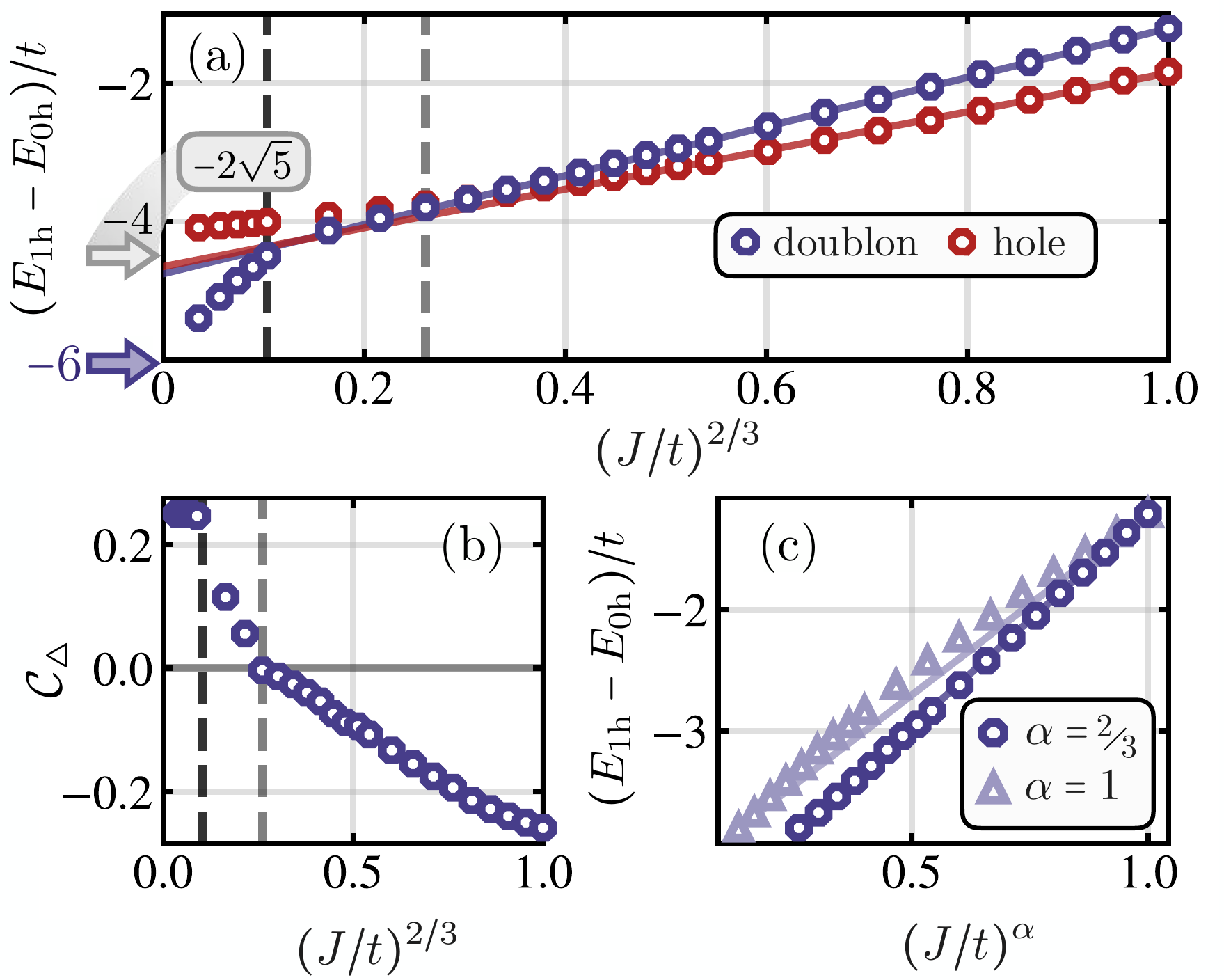}
\caption{(a) Ground state hole and doublon energies in the singly doped triangular $t-J$ model as a function of $(J/t)^{2/3}$. $E_{1h}$ ($E_{0h}$) denotes the energy of the singly doped (undoped) system. For intermediate $J/t$, both ground state energy curves show linear behavior with extrapolated $y$-intercept close to $-2\sqrt{5}$ (grey arrow), as predicted by geometric string theory, Eq.~\eqref{eq:gst_E_scaling}. At small values of $J/t$, the formation of the Nagaoka polaron is observed for doublons by a kink in the energy, which is now approaching $-6t$ for $J/t \rightarrow 0$ (blue arrow). The transition is underlined in (b), where $\mathcal{C}_{\triangle}$ changes sign at around $(J/t)^{2/3} \sim 0.25$ (light grey dashed lines) followed by a sudden jump to strong FM signals at $(J/t)^{2/3} \sim 0.1$ (dark grey dashed lines). (c) For doublons the power-law scaling $\sim (J/t)^{2/3}$ is corroborated, where the ground state energies are shown as a function of $(J/t)^{\alpha}$ for both $\alpha=1$ and $\alpha = \nicefrac{2}{3}$. In the former case, curvature is clearly visible, indicated by the solid light blue line connecting the first and last data point in the plot. For $\alpha = \nicefrac{2}{3}$, the curvature disappears. }
\label{fig:gs}
\end{figure}

In order to test the prediction Eq.~\eqref{eq:gst_E_scaling}, we calculate ground state energies of a single dopant as a function of $(J/t)^{2/3}$, presented in Fig.~\ref{fig:gs}~(a). For $(J/t)^{2/3}\gtrsim 0.2$, both hole and doublon energies seem to be well approximated by a linear scaling law. Fig.~\ref{fig:gs}~(c) corroborates the scaling of the doublon energy by plotting the data against $(J/t)^{\alpha}$, for both $\alpha = 1, \nicefrac{2}{3}$. In the case of $\alpha = 1$ (light blue triangles), positive curvature clearly remains visible, as underlined by the solid line connecting the first and last data point in the plot. For $\alpha=\nicefrac{2}{3}$, in turn, the curvature almost completely vanishes (dark blue dots). Indeed, when treating $\alpha$ as an independent fit parameter, we find $\alpha = 0.67$, further stressing the predicted $\nicefrac{2}{3}$ scaling of the doublon energy.

In the case of hole doping, our data is more consistent with $\alpha=1$~\cite{supp_mat}. We speculate that terms $\sim (J/t)^1$ in Eq.~\eqref{eq:gst_E_scaling} may contribute stronger to the signal; finite-size effects may also play a role for understanding the observed scaling. A universal scaling of the form Eq.~\eqref{eq:gst_E_scaling} can hence not be ruled out in the thermodynamic limit based on our currently achievable numerics.

Importantly, for both doublon and hole, extrapolated linear fits as a function of $(J/t)^{2/3}$ to $J/t \rightarrow 0$ show almost perfect match with the predicted asymptotic value from geometric string theory, only slightly overshooting $E_0/t = -2\sqrt{z-1}$. Indeed, finite size scaling for widths $L_y = 3,6$ suggests corrections towards $E_0/t$ in the thermodynamic limit~\cite{supp_mat}.

Furthermore, in the relevant regime for geometric strings, $0.2 \lesssim (J/t)^{2/3} \lesssim 1.0$, we observe that both doublon and hole energies match up to only minor discrepancies. This, in turn, corroborates linear confining behavior also for hole doped systems, as in first approximation the string picture is independent of the hopping sign and only relies on locally present AFM correlations around the dopant. For growing $J/t$, however, where the dispersion of spinons becomes more prominent~\cite{Grusdt_strings, Zhu2022}, corrections are expected from the scaling behavior that do depend on the dopant species, as also seen here for $(J/t)^{2/3} \gtrsim 0.5$.

In the doublon doped system, the onset of Nagaoka FM for $J/t \rightarrow 0$ reveals itself by a clearly visible kink of the doublon energy away from the string picture scaling at around $(J/t)^{2/3} = 0.1$ ($J/t \sim 30$), see the dark dashed line in Fig.~\ref{fig:gs}~(a). Once the Nagaoka polaron builds up around the doublon, it can gain kinetic energy that, in case of a fully polarized background for $J/t \rightarrow 0$, approaches the value of a free dopant, $E_{\text{kin}} = -6t = -zt$. 

For the hole doped system, in contrast, the minimal kinetic energy in a fully polarized spin background is given by $E_{\text{kin}} = -3t>-zt$, i.e., it is kinetically frustrated. This limit lies above the asymptotic value $-2\sqrt{5} t \approx -4.47 t$ expected from the string picture. This is in agreement with our numerics, which confirm hole energies $< -3t$ for $(J/t)^{2/3}\lesssim 0.6$ (where AFM correlations are present around the dopant), demonstrating that the hole can gain more kinetic energy by instead moving in an AFM spin environment. For $J/t \rightarrow 0$, the hole's energy reaches roughly $-4.1 t$, supplementing numerical predictions presented in~\cite{Haerter2005, Sposetti2014}.

Fig.~\ref{fig:gs}~(b) shows charge-spin-spin correlations $\mathcal{C}_{\triangle}$ for the doublon doped system. For decreasing $J/t$, we observe that $\mathcal{C}_{\triangle}$ changes sign at around $(J/t)^{2/3} \sim 0.25$ ($t/J \sim 8$), before suddenly jumping to strongly positive FM correlations at $(J/t)^{2/3} \sim 0.1$ ($J/t \sim 30$) -- coinciding with notable deviations from the predicted energy scaling, see the dark and light grey dashed lines in Fig.~\ref{fig:gs}~(a) and (b). Pictorially, a Nagaoka bubble of positive correlations around the hole forms, which grows with decreasing $J/t$ until the fully polarized state is achieved (the radius of the Nagaoka bubble roughly scales as $\propto (t/J)^{1/4}$, as can be seen from variational arguments~\cite{White2001}). In the thermodynamic limit, the local correlator $\mathcal{C}_{\triangle}$ is expected to feature a similar behavior with an observable sign change of correlations in direct vicinity of the dopant at intermediate $J/t$, smoothly evolving towards the Nagaoka phase at $J \rightarrow 0$, whereby the total spin of the system $S$ continuously grows from $S=0$ to $S= S_{\text{max}}$ in a series of phase transitions~\cite{supp_mat}.

\textit{Outlook.---} We have analyzed the competing interplay between kinetic and magnetic frustration in the singly doped triangular $t-J$ model, associated with energy scales $t$ and $J$, respectively, and at temperature $T$. We unveiled regimes of itinerant magnetism in the vicinity of the dopant, which can directly be accessed and observed using state-of-the-art quantum gas microscopy of ultracold atoms on triangular optical lattices. Remarkably, we predict an increased stability of Nagaoka ferromagnetism at elevated temperatures, potentially facilitating its observation with ultracold atoms. Studying Nagaoka FM at finite doping as well as on other non-bipartite lattices are interesting directions for further research~\cite{Hanisch1995}.

At low temperatures, we have presented strong evidence for the existence of linear confinement of the dopants surrounded by $120^{\circ}$ ordered spins on the triangular lattice. The strikingly accurate scaling of the doublon's energy $\sim (J/t)^{2/3}$ firmly establishes the picture of geometric strings for frustrated systems. The role of enhanced quantum fluctuations as well as the non-othogonality of classically ordered $120^{\circ}$ spin states on the geometric string basis construction will be discussed in future work. Particularly, our work further advocates a possible significance of strings for pairing~\cite{Grusdt2022, Hirthe2022} and ro-vibrational excitations~\cite{GrusdtX} in doped AFMs -- hinting towards an unprecedented universal picture of the physics in doped Mott insulators.

\textit{Note added.} During the preparation of this manuscript, spin polarons have been directly imaged in triangular lattice ultracold atom experiments in \cite{prichard2023, lebrat2023}, being in support of the theoretically predicted results in this manuscript. Furthermore, Nagaoka ferromagnetism has been analyzed in modified Fermi-Hubbard models on square and triangular lattice geometries in~\cite{samajdar2023}.

\textit{Acknowledgments.---} We thank E. Demler, M. Greiner, K. Hazzard, A. Kale, L. Kendrick, M. Lebrat, I. Morera, and M. Xu for fruitful discussions. This research was funded by the Deutsche Forschungsgemeinschaft (DFG, German Research Foundation) under Germany’s Excellence Strategy—EXC-2111—390814868, by the European Research Council (ERC) under the European Union’s Horizon 2020 research and innovation programme (grant agreement number 948141), and by the NSF through a grant for the Institute for Theoretical Atomic, Molecular, and Optical Physics at Harvard University and the Smithsonian Astrophysical Observatory.

\widetext
\appendix
\section{\underline{Supplementary Materials}}

\subsection{Finite temperature calculations}

We use purification schemes within the MPS framework to simulate the $t-J$ model on the triangular lattice at finite temperature. We conserve the system's $\rm{U(1)}$ charge conservation symmetry, while employing a grand-canonical ensemble in the spin sector that allows for finite thermal spin magnetizations.

We here shortly summarize the purification schemes used in the main text, and refer to e.g.~\cite{Nocera2016, Schloemer2022, Schloemer2022_recon} for more detailed discussions. When purifying the system, the Hilbert space is enhanced by one auxiliary (often also called ancilla) site per physical site, which allows to display mixed states in the physical subset of the Hilbert space as pure states on the enlarged space. Thermal expectation values are then computed via
\begin{equation}
    \braket{\hat{O}}_{\beta} = \frac{\braket{\Psi(\beta)|\hat{O}|\Psi(\beta)}}{\braket{\Psi(\beta)|\Psi(\beta)}},
\end{equation}
where $\ket{\Psi(\beta)} = e^{-\beta \hat{H}/2} \ket{\Psi(\beta=0)}$ is the maximally entangled state $\ket{\Psi(\beta=0)}$ evolved in imaginary time $\tau = \beta/2$, and $\mathcal{O}$ is an operator acting on the physical sites only (i.e., all auxiliary degrees of freedom are traced out in the evaluation of the expectation value). Note that this is indeed an exact formulation of the usual form $\braket{\hat{\mathcal{O}}} = \frac{1}{Z} \text{Tr} ( \rho \hat{O} )$, where $Z = \text{dim}(\mathcal{H}) \braket{\Psi(\beta)|\Psi(\beta)}$ with $\text{dim} (\mathcal{H})$ the dimension of the Hilbert space. 

The maximally entangled state for the triangular $t-J$ model at infinite temperature and in the canonical (grand-canonical) ensemble for the charge (spin) sector reads
\begin{equation}
\label{eq:max_ent}
    \ket{\Psi(\beta=0)} = \hat{\mathcal{P}}_{N} \bigotimes_{i = 0}^{N_L-1} \left( \ket{0,0} + \sum_{\sigma=\uparrow,\downarrow} \ket{\sigma, \bar{\sigma}} \right).
\end{equation}
Here, $N_L= L_x L_y$ the total number of physical sites in the ladder system, $\{\ket{0}, \ket{\uparrow}, \ket{\downarrow}\}$ is the single particle basis of the $t-J$ model with $\bar{\uparrow} = \downarrow$, $\bar{\downarrow} = \uparrow$, and $\hat{\mathcal{P}}_{N}$ is the projector to the subspace with $N$ particles in the physical system (in our case fixed to $N = N_L-1$); the first and second entries in the kets correspond to physical and auxiliary sites, respectively. \\

In order to get the MPS representation of the maximally entangled state, we perform a ground state search of specifically tailored entangler Hamiltonians, cf.~\cite{Nocera2016}. After retrieving the maximally entangled state, we employ imaginary time evolution techniques to evolve the state away from $\beta=0$ towards finite temperatures.

Since the infinite temperature states (being projected product states) are of low bond dimension, local approximations of the Hamiltonian (and subsequent exponentiation) will suffer from large projection errors and are of low quality. Hence, we start by employing global methods for a single step in imaginary time, after which the entanglement in the system (and the bond dimension of the thermal MPS) has sufficiently increased to switch to local methods. 

In particular, we start with two global Krylov steps with $\Delta \tau = 0.01$, after which we switch to the local two-site TDVP method~\cite{Paeckel_time} with $\Delta \tau = 0.02$. At $\tau = 2.0$, we increase the imaginary time step to $\Delta \tau = 0.1$ in order to evolve the system down to low temperatures and compare to ground state properties. For our calculations, we fix weight cutoffs to $w_{\text{Kry}} = 10^{-7}$ and $w_{\text{TDVP}} = 10^{-9}$ for the Krylov and TDVP imaginary time evolution, respectively, and limit maximum bond dimensions to $\chi_{\text{Kry}} = 1024$, $\chi_{\text{TDVP}} = 3000$ and $\chi_{\text{TDVP}} = 5000$ to test convergence. 

\begin{figure}
\centering
\includegraphics[width=\textwidth]{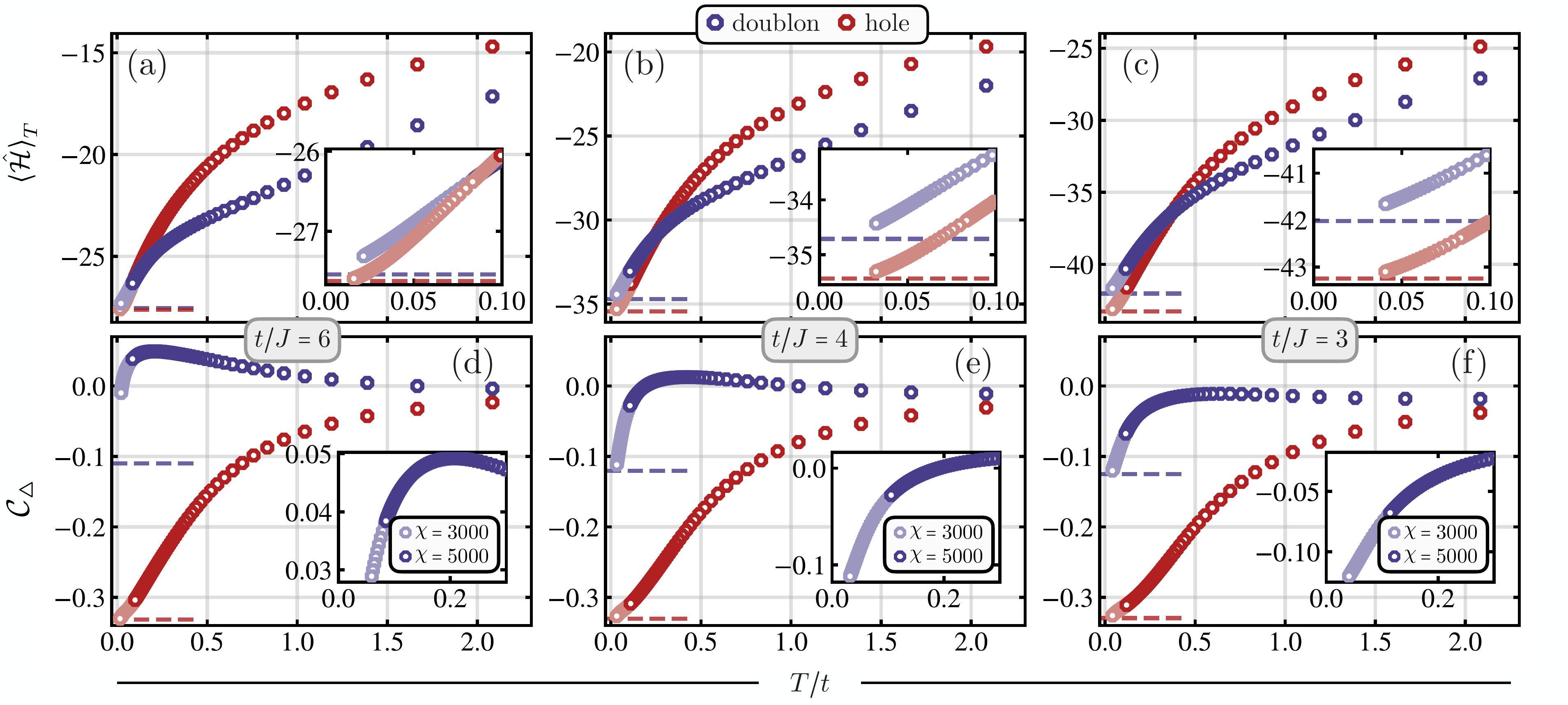}
\caption{\textbf{Convergence of finite temperature DMRG.} (a)-(c) Thermally averaged energy $\braket{\hat{\mathcal{H}}}_T$ as a function of temperature $T/t$ for (a) $t/J = 6$, (b) $t/J = 4$, and (c) $t/J = 3$, for both a doped hole (red) and doublon (blue). In all cases, convergence towards the ground state energies (dashed red and blue lines) can be observed, underlined by the zoom-in panels showing the low temperature regime. (d)-(f) Local charge-spin-spin correlator $\mathcal{C}_{\triangle}$ as a function of $T/t$, evaluated for $\mathbf{i} = [x=4,y=2]$ in the physical system. Down to $T/t \sim 0.1$, convergence of maximal bond dimensions $\chi = 3000$ and $\chi=5000$ is observed, as illustrated by the insets.}
\label{fig:conv}
\end{figure}

Convergence of the energy as a function of temperature $T/t$ is shown for the $9 \times 3$ systems in Fig.~\ref{fig:conv}~(a)-(c). Energies are seen to converge towards the ground state results (dashed lines) both for the hole (red data) and doublon (blue data) doped systems. The insets are zoom-in panels showing the low-temperature region $T/t = 0 \dots 0.1$, where in particular for $t/J = 3$ convergence towards the ground state can be clearly established. 

Fig.~\ref{fig:conv}~(d)-(f) show spin-spin correlations around the dopant, $\mathcal{C}_{\triangle}$, as a function of temperature, cf. Fig.~\ref{fig:fT} in the main text, where again convergence towards the ground state results can be observed. As discussed in the main text, capturing the turnover from increasingly FM signals caused by kinetic effects to strong AFM correlations on the doublon doped side is a difficult task. In the insets of Fig.~\ref{fig:conv}~(d)-(f), we show zoom-ins into the peak region -- showing convergence down to $T/t \sim 0.1$. 

\subsection{Hole energy scaling}

\begin{figure}
\centering
\includegraphics[width=0.65\textwidth]{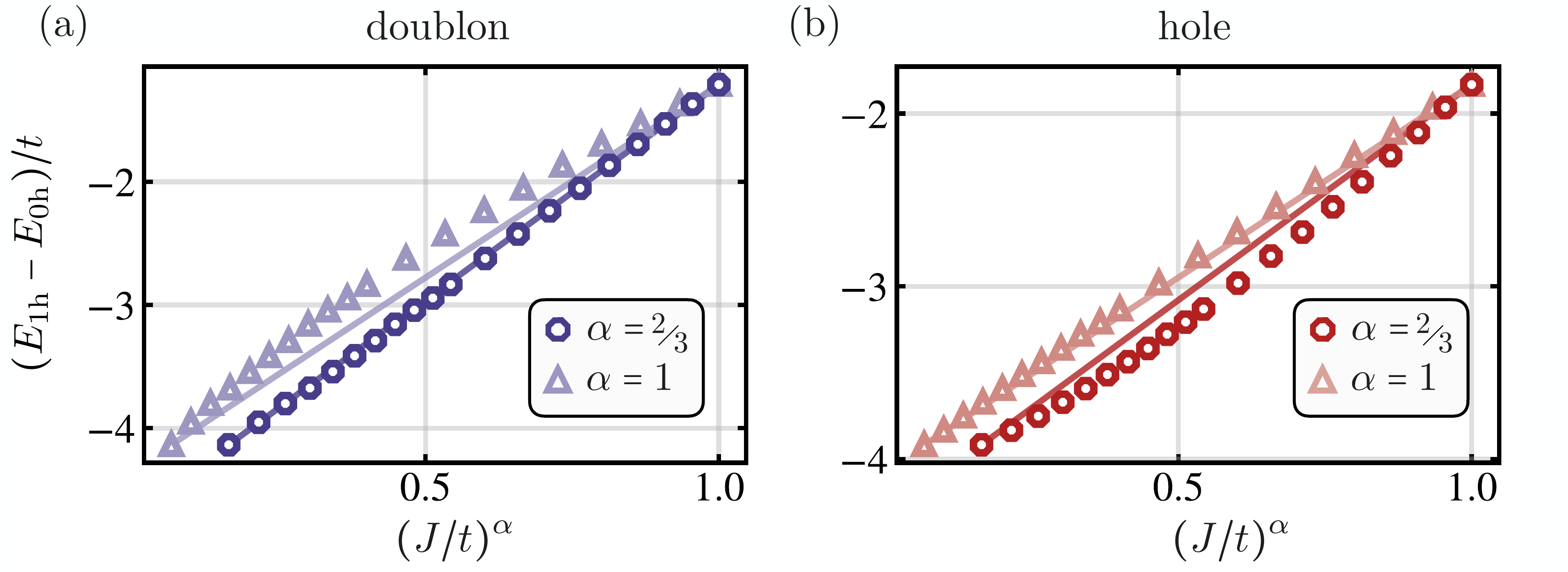}
\caption{\textbf{Single dopant energy scaling.} Energies of a single doublon (hole) as a function of $(J/t)^{\alpha}$ for $\alpha=1$, $\alpha=\nicefrac{2}{3}$ for a doped doublon, (a), and hole, (b). As shown in the main text, the energy scaling on the doublon doped side matches the string prediction, Eq.~\eqref{eq:gst_E_scaling_supp} to remarkable precision. For hole doping, our results are more consistent with $\alpha = 1$, though a scaling of the form Eq.~\eqref{eq:gst_E_scaling_supp} can not be ruled out in the thermodynamic limit.}
\label{fig:Escaling_supp}
\end{figure}

In the main text, we have established clear evidence for the existence of linear confining potentials in the doublon doped case by analyzing the dopants energy scaling as a function of $(J/t)^{\alpha}$. The vanishing curvature for $\alpha = \nicefrac{2}{3}$ constitutes strong affirmation for the applicability of the string picture, which predicts a scaling of the form  
\begin{equation}
    E/t = -2\sqrt{z-1} + const. \, (J/t)^{2/3} + \mathcal{O}(J/t),
    \label{eq:gst_E_scaling_supp}
\end{equation}
for $J/t \ll 1$. Results for the doublon doped system are shown in Fig.~\ref{fig:Escaling_supp}~(a), cf. Fig.~\ref{fig:gs}~(c) in the main text.
The hole doped case is shown in Fig.~\ref{fig:Escaling_supp}~(b). The data for the hole doped system is seen to be more consistent with $\alpha=1$, whereby the energies plotted against $(J/t)^{2/3}$ shows noticeable negative curvature. As argued in the main text, we speculate that higher order terms $\mathcal{O}(J/t)$ in Eq.~\eqref{eq:gst_E_scaling_supp} may contribute stronger to the hole's energy, making it more difficult to isolate a possible $\alpha = \nicefrac{2}{3}$ contribution. Finite size effects might further lead to pronounced influences on the observed energies, such that we can not rule out a scaling of the form Eq.~\eqref{eq:gst_E_scaling_supp} in the thermodynamic limit. For instance, the close proximity of the y-intercept assuming a scaling of the form $\alpha = \nicefrac{2}{3}$ to the theoretically predicted value $-2\sqrt{5}$ as well as the almost indistinguishable energies of holes and doublons in the regime $(J/t)^{2/3} \sim 0.2 \dots 0.4$ are in strong support of major characteristics of the string theory prediction, Eq.~\eqref{eq:gst_E_scaling_supp}.

\subsection{Finite size effects}

In the main text, systems of widths $L_y = 6$ were discussed for the ground state calculations. By comparing to systems of width $L_y = 3$, we here analyze trends when going towards the thermodynamic limit. 

\begin{figure}
\centering
\includegraphics[width=0.3\textwidth]{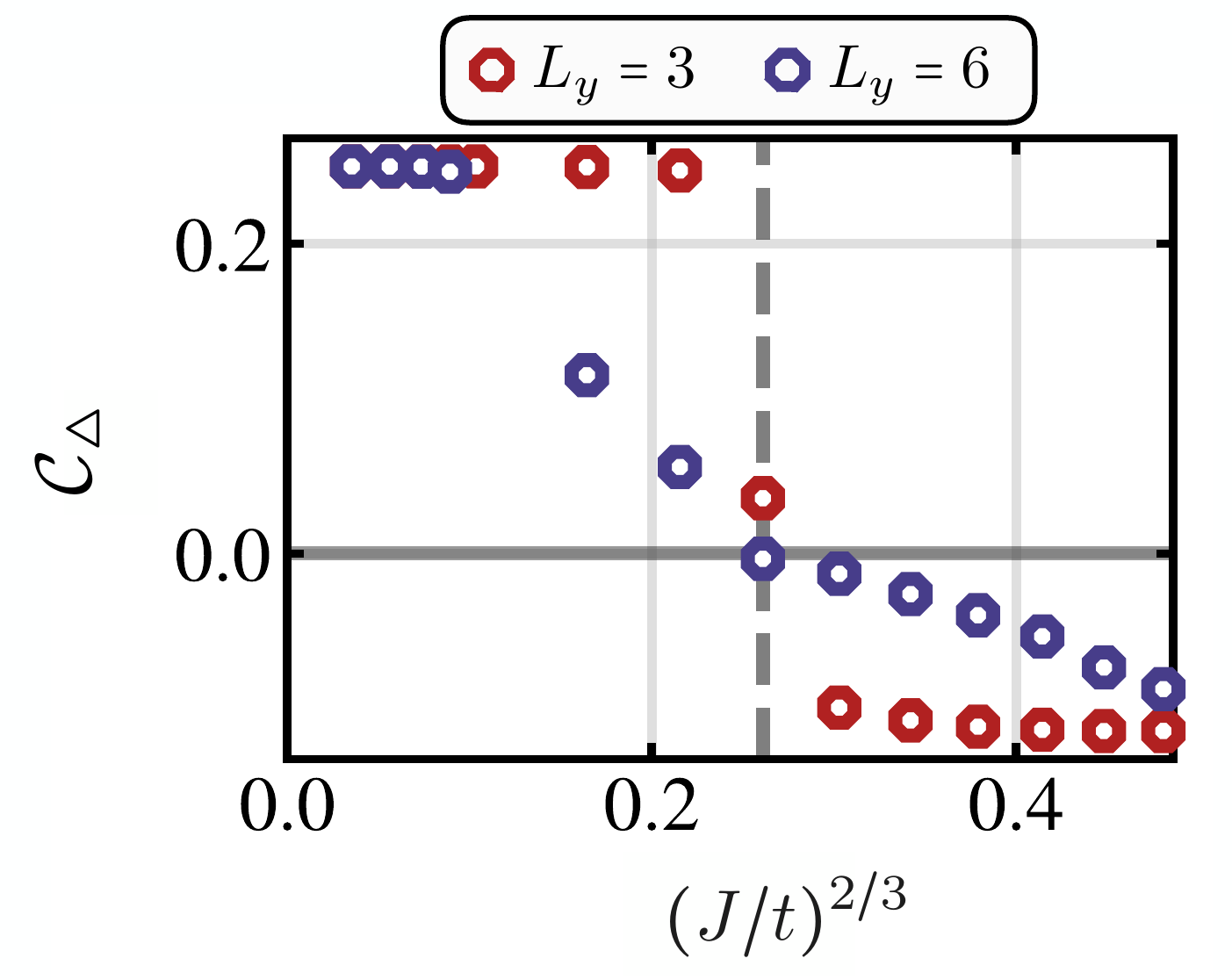}
\caption{\textbf{Finite size effects.} Charge-spin-spin correlations $\mathcal{C}_{\triangle}$ for systems of sizes $L_x \times L_y = 9 \times 3$ (red) and $L_x \times L_y = 6 \times 12$ (blue), evaluated for $\mathbf{i} = [x=4,y=2]$ and $\mathbf{i} = [x=4,y=3]$, respectively. The dashed line indicates the sign changes of both systems, coinciding at $(J/t)^{2/3}\sim 0.25$, cf. Fig.~\ref{fig:gs}~(b) in the main text.}
\label{fig:fse}
\end{figure}

In Fig.~\ref{fig:fse}, we show the charge-spin-spin correlator $\mathcal{C}_{\triangle}$ as a function of $(J/t)^{2/3}$, cf. Fig.~\ref{fig:gs}~(b) in the main text. We see that the sign change of $\mathcal{C}_{\triangle}$ coincides both for $L_y = 3$ as well as $L_y = 6$ at approximately $(J/t)^{2/3}\sim 0.25$ (corresponding to $t/J \sim 8$). Pictorially, a Nagaoka bubble of positive correlations around the hole forms, which grows with decreasing $J/t$ until the fully polarized state is reached (the radius of the Nagaoka bubble roughly scales as $\propto (t/J)^{1/4}$, as can be seen from variational arguments~\cite{White2001}). For finite systems, this happens at a finite value of $J/t$, with decreasing $J/t$ as the system size is increased. This suggests that, in the thermodynamic limit, the local correlator $\mathcal{C}_{\triangle}$ features a similar behavior with an observable sign change at intermediate $J/t$, smoothly evolving towards the saturated Nagaoka phase at $J \rightarrow 0$ with $\mathcal{C}_{\triangle} = \nicefrac{1}{4}$, whereby the total spin of the system $S$ continuously grows from $S=0$ to $S= S_{\text{max}}$ in a series of phase transitions.

\begin{figure}
\centering
\includegraphics[width=0.9\textwidth]{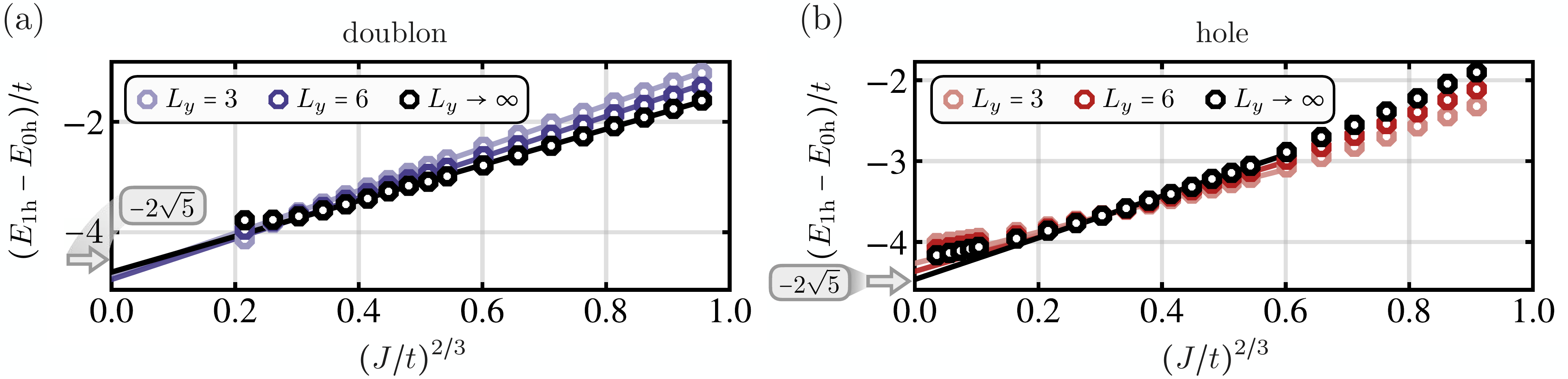}
\caption{\textbf{Finite size scaling.} Doublon, (a), and hole, (b), energies as a function of $(J/t)^{2/3}$ and for system sizes $L_x \times L_y = 9 \times 3$ (light colors) and $L_x \times L_y = 12 \times 6$ (dark colors). Extrapolations to infinite system sizes are shown by black data points. Linear fits show corrections towards the universal prediction of $-2\sqrt{5}$ for $J/t \rightarrow 0$. Linear fits are done for $(J/t)^{2/3}>0.3$ for doublons and $0.3<(J/t)^{2/3}<0.6$ for holes.}
\label{fig:fss}
\end{figure}

Shown in Fig.~\ref{fig:fss} are the results of the doublon and hole energies in Fig.~\ref{fig:fss}~(a) and (b), respectively, as a function of $(J/t)^{2/3}$ and for systems sizes $L_x \times L_y = 9 \times 3$ (light colors) and $L_x \times L_y = 12 \times 6$ (dark colors). Extrapolations to infinite system sizes by a $1/L_y$ expansion are shown by black data points. We note that the extrapolation to infinite system sizes by two cylinder widths is a very crude procedure, and must be interpreted with lots of caution. Though a slight negative curvature trend is visible in the extrapolation of the doublon energies, the presence of the Nagaoka regime at $J/t \rightarrow 0$ suggests a dip towards energies $-6t$ also in the thermodynamic limit (we do not extrapolate data below $(J/t)^{2/3}$, cf. Fig.~\ref{fig:gs}~(b) in the main text). This trend is, however, not visible in the extrapolation, and hence we attribute the slight positive curvature to the rudimentary finite size extrapolation. In the thermodynamic limit, we hence do not rule out a linear scaling of the doublon's energy as a function of $(J/t)^{2/3}$. 

Though the data for the doped hole is more consistent with a linear scaling, we plot the hole's energy as a function of $(J/t)^{2/3}$ in Fig.~\ref{fig:fss}~(b) in order to identify possible trends when increasing the system's width. Though a slight negative curvature is still observable in the extrapolated thermodynamic limit, we nevertheless see corrections of the y-intercept of fitted linear curves in the approximatelty linear regime towards the universal prediction in Eq.~\eqref{eq:gst_E_scaling_supp}. However, more detailed work and larger system sizes are needed to distill the true scaling behavior of the doped hole.

\subsection{Density profiles}

\begin{figure}
\centering
\includegraphics[width=0.7\textwidth]{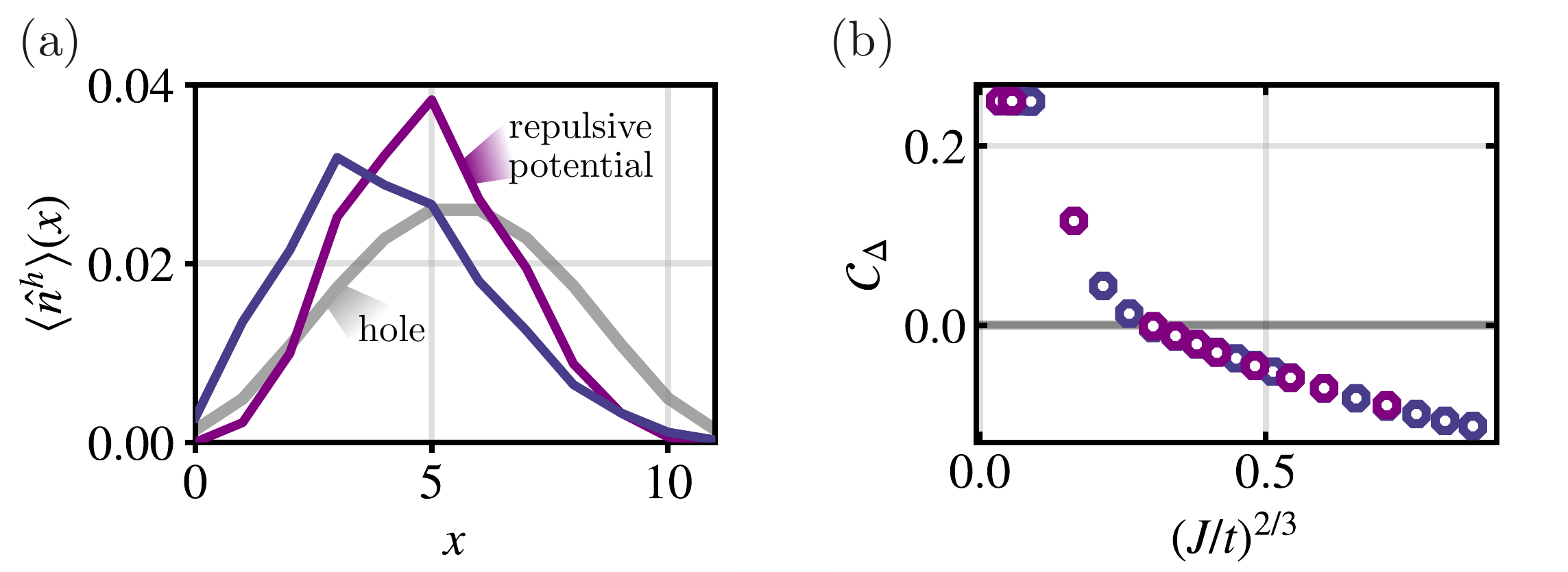}
\caption{\textbf{Density profiles and hole-spin-spin correlations.} (a) Density profiles for the single hole (grey) and doublon (blue) for the $L_x \times L_y = 6 \times 12$ system, at $J/t = 0.2$ along $y=2$. The doublon's density distribution is observed to be asymmetric around the center of the system. We apply strong repulsive potentials at the open edges along $x$ of the system (with strength $100 J$), which shifts the distribution closer to the center (though a slight asymmetry is still observed). Spin-spin correlations around the dopant at the maximum of the distribution is shown with and without repulsive potentials in blue and purple, respectively, in (b) -- which are seen to  remain unchanged in the two cases.}
\label{fig:conv_pot}
\end{figure}

Fig.~\ref{fig:conv_pot}~(a) shows the dopant's density distribution both for the doped hole (grey) as well as doublon (blue) for the $L_x \times L_y = 6 \times 12$ system, for $J/t = 0.2$. Shown are cuts at constant $y$ (where all values of $y$ are equivalent due to applied periodic boundary conditions). In contrast to the doped hole, the doublon's density distribution is seen to be asymmetric around the center of the system, which does not significantly change upon increasing the bond dimension further (shown are results with $\chi = 5500$). We apply strong repulsive potentials for the hole density along the open edges of the system (with strength $100 J$), which shifts the distribution closer to the center, see the purple curve in Fig.~\ref{fig:conv_pot}~(a). However, spin correlations around the dopant remain unchanged, as shown in Fig.~\ref{fig:conv_pot}~(b), where we choose the reference site $\mathbf{i}_0$ to coincide with the maximum doublon density along $x$. We note that for systems of width $L_y = 3$, the doublon's density does converge to a symmetric distribution, see Fig.~\ref{fig:qpw}~(b) -- underlining that there is no qualitative change expected from the asymmetry of the doublon's density for $L_y = 6$.

\subsection{Odd-even effects of $\mathcal{C}_{\triangle}$}

\begin{figure}
\centering
\includegraphics[width=0.45\textwidth]{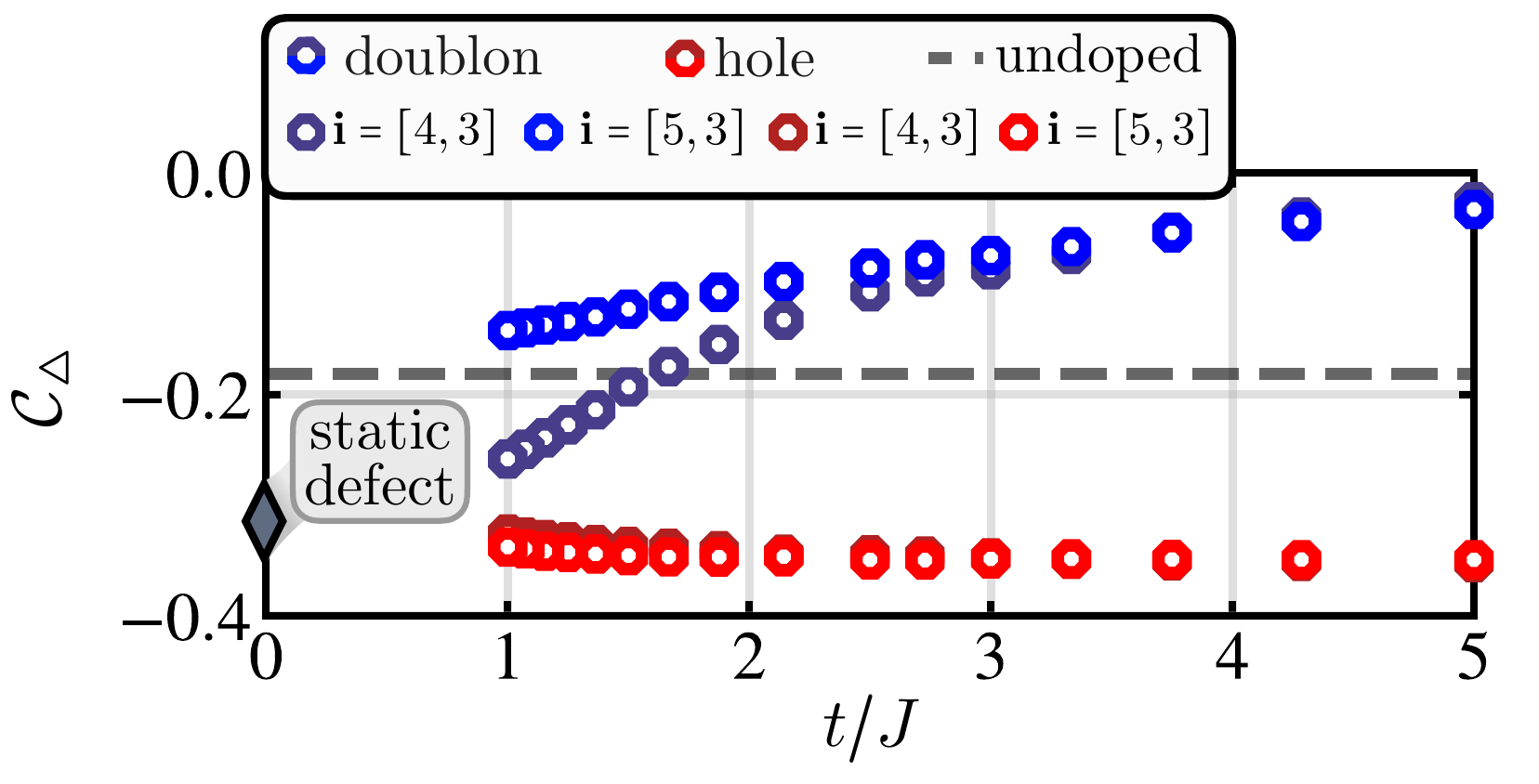}
\caption{\textbf{Odd-even effects.} Charge-spin-spin correlations, $\mathcal{C}_{\triangle}$ for doublon (blue colors) and hole (red colors) doping, on two triangular plaquettes with $\mathbf{i} = [x=4, y=3]$ (dark colors) and $\mathbf{i} = [x=5, y=3]$ (light colors). Notable effects are visible for $t/J \lesssim 2.5$, presumably stemming from the broken transnational symmetry in the system.}
\label{fig:oee}
\end{figure}

For the charge-spin-spin correlator $\mathcal{C}_{\triangle}$, we take a look at its dependence on the position of the triangular plaquette along the long side of the cylinder. Results for $\mathbf{i}= [x=4,y=3]$ (as shown in the main text) and $\mathbf{i} = [x=5,y=3]$ are shown in Fig.~\ref{fig:oee}. For $t/J \lesssim 2.5$, notable odd-even modulations are visible in the correlator. We believe that these modulations of the observable along the long direction originate from the broken transnational symmetry, which mixes the degenerate ground states existing at dispersion minima in the thermodynamic limit. 

\subsection{Quasiparticle spectral weight and self-localization}

In~\cite{Zhu2022}, it has been demonstrated that the quasiparticle properties of the magnetic polarons in the hole and doublon doped triangular lattice significantly differ from one another. In the hole doped case, the quasiparticle was found to be dispersive, while the magnetic polaron in the doublon doped case is nearly flat with strongly suppressed quasiparticle weight $Z(\mathbf{k})$. We calculate $Z_{\text{gs}}$ for the ground state as a function of $J/t$, which can be accessed through the DMRG MPS ground states by evaluation of
\begin{equation}
    Z_{\text{gs}} = \sum_{\sigma} \sum_{\mathbf{i}} |\braket{\psi_{1h}|\hat{c}_{\mathbf{i}, \sigma}|\psi_{0h}}|^2,
\end{equation}
where $\ket{\psi_{1h}}$ ($\ket{\psi_{0h}}$) denote the ground state with a single and no hole, respectively. 

\begin{figure}
\centering
\includegraphics[width=0.75\textwidth]{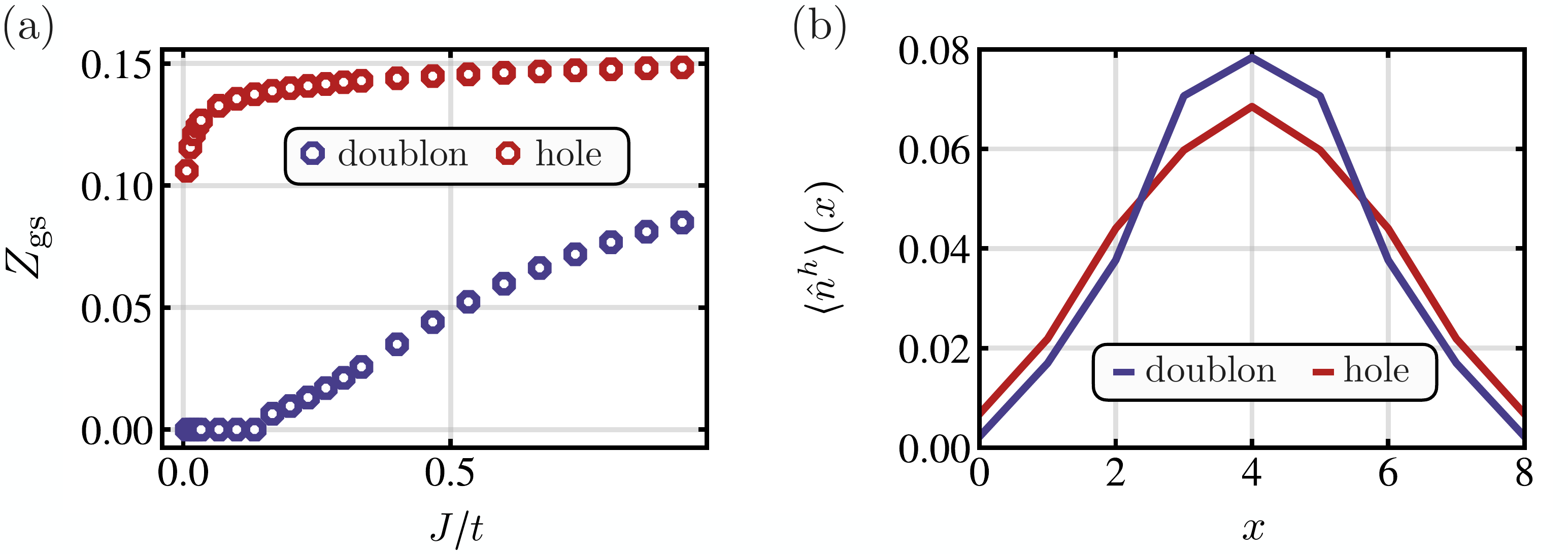}
\caption{\textbf{Quasiparticle weight and hole density} (a) Ground state quasiparticle weight $Z_{\text{gs}}$ as a function of $J/t$ for a single doped hole (red) and doublon (blue). For doublons, the Nagaoka regime leads to strong suppression of the quasiparticle weight, suggesting a flat quasiparticle dispersion. (b) This is underlined by the hole density $\braket{\hat{n}^h}(x)$, showing enhanced localization of the doublon compared to the hole, here shown for $J/t = 0.2$.}
\label{fig:qpw}
\end{figure}

Results are shown in Fig.~\ref{fig:qpw}~(a). In the hole doped case, strong AFM correlations (cf. Fig.~\ref{fig:gs_corr} in the main text) suggest a large quasiparticle weight, as confirmed by our numerics. For doublons, on the other hand, the quasiparticle weight is strongly suppressed at low $J/t$, where the build up of Nagaoka FM leads to a vanishing overlap between $\ket{\psi_{1h}}$ and $\hat{c}_{\mathbf{i}, \sigma}\ket{\psi_{0h}}$. 

The hole density distribution, $\braket{\hat{n}^h}(x)$, is shown for $J/t = 0.2$ in Fig.~\ref{fig:qpw}~(b). In the case of doublon doping, the density is noticeably more localized compared to hole doping, corroborating the picture of a heavy Nagaoka magnetic polaron on the doublon doped side and supplementing results presented in~\cite{Zhu2022}.   

\twocolumngrid
\bibliography{triangular}
\onecolumngrid
\pagebreak
\widetext

\end{document}